\documentstyle[12pt,psfig,aps]{revtex}

\begin{document}
 
\title{Unlike Particle Correlations and the Strange Quark Matter Distillation 
Process}
\author{
        D.~Ardouin\footnote
{on leave from University of Nantes, U.M.R. Subatech},
        Sven~Soff, C.~Spieles, S.~A.~Bass, H.~St\"ocker}
        
\address{
        Institut f\"ur Theoretische Physik, J. W. Goethe-Universit\"at,\\
        Postfach 11 19 32, D-60054 Frankfurt am Main, Germany;
\footnote{supported by GSI, BMBF, DFG and Buchmann Fellowship}}

\author{D.~Gourio, S.~Schramm}
\address{
        GSI Darmstadt, Postfach 11 05 52, D-64220 Darmstadt, Germany;}
        
\author{C.~Greiner}
\address{Institut f\"ur Theoretische Physik, J. Liebig-Universit\"at,\\
        Heinrich-Buff-Ring 16, D-35392 Gie{\ss}en, Germany;}
        
\author{R.~Lednicky\footnote{supported by GA AV CR, Grant No. A1010601 
and GA CR, Grant No. 202/98/1283}}  
\address{Institute of Physics of the Academy of Sciences
of the Czech Republic,\\
        Na Slovance 2, 18040 Prague 8, Czech Republic;}
        
\author{V.~L.~Lyuboshitz\footnote{supported by RFFI, Grant No.97-02-16699}}
\address{JINR Dubna, 141980, Moscow, Russia;}

\author{J.-P.~Coffin, C.~Kuhn}
\address{CRN Strasbourg, Universit\'e L. Pasteur, Strasbourg, France.}

\maketitle

\begin{abstract}
We present a new technique for observing the strange quark matter distillation 
process based on unlike particle correlations. A simulation is 
presented based on the scenario of a two-phase thermodynamical evolution model.
\end{abstract}

\pacs{25.75.+r,12.38.Mh,12.39.Ba,12.39.Mh}

\section{Motivation} 

The possibility to create strangelets or droplets of metastable cold
strange quark matter in ultra-relativistic collisions has been
proposed and studied by several authors 
\cite{Witten84,FarhiJaffe84,LiuShaw84,CGreiner87,CGreiner88,LeeHeinz89,Barz88}.
The existence of such exotic states, as well as metastable exotic
multi-strange baryonic objects (MEMO's), has fundamental importance in
cosmological models and in the underlying description of strong
interactions.
Several experiments are being carried out at the Brookhaven AGS (E864, E878) and the  
CERN SPS (NA52), which look for strangelet production (searching for small  $Z/A$ ratios) 
\cite{NA52,E864,E878}.

Among the different theoretical approaches 
\cite{Witten84,FarhiJaffe84,LiuShaw84,CGreiner87,CGreiner88,LeeHeinz89,Barz88}, 
a mechanism of separation of strangeness from anti-strangeness 
(distillation process) has been proposed \cite{CGreiner87} 
during hadronization of a system at finite baryon densities. 
This scenario, which assumes a first order phase transition, 
predicts a relative time delay between the production of 
strange and anti-strange particles. 
In the present paper, we propose 
to search for this delay with the help of the novel
correlation method proposed in \cite{Led9422,Led96}. It exploits the
sensitivity of the directional dependence of the correlation function
of two non-identical particles to the time order of their
emission. This new technique can allow for 
the study of the transient strange quark 
matter state even if it
decays on  strong interaction time scales \cite{Era95,Era96,Soff97a}.

A different method, exploiting $K_s^0K_s^0$ correlations, also
allowing for a study of the relative delays (but not their sign) 
in $K$ and $\bar{K}$
emission was proposed in \cite{Gyulassy92}. 
It is based on the fact that
the $K_s^0K_s^0$ system, due to the positive CP parity of $K_s^0$, 
triggers out the symmetric combination of the $K^0\bar{K}^0$ states, thus leading to the familiar
Bose--Einstein correlation pattern \cite{Lyubo79}.
This pattern, contrary to the case of non interacting identical bosons, 
is however substantially modified by the effect 
of the strong final state interactions \cite{Lednicky82}.

Here we will use the general method of \cite{Led9422,Led96} to study the
impact of the time delays on the correlations in $K^+K^-$ pairs.
The choice of this system, similar to the case of neutral kaons, 
apart from the possibility to determine the signs of the time delays, 
has the advantage of weaker distortions caused by the resonance
production (more than 50\% of kaons are predicted to originate from
direct emission \cite{Spieles97}), as compared to pions, 
while the experimental feasibility is better than in the
case of neutral kaons.

\section{The mixed phase thermodynamical approach}

The  dynamical evolution of the mixed phase consisting of a quark 
gluon plasma and hadronic gas will be described in a 
two-phase model which takes into account equilibrium 
as well as non-equilibrium features \cite{CGreiner91,Spieles96}. 

Within this model, two main assumptions are made. 
Firstly, the QGP is surrounded by a layer of hadron gas and 
equilibrium ( described by Gibbs conditions) is  maintained during the
evolution. Secondly, non-equilibrium evaporation is incorporated by a time
dependent emission of hadrons from the surface of
the hadronic fireball. 
Within this model, it is possible to follow the 
evolution of mass, entropy and strangeness fraction $f_s$ of  
the system and extract relative yields of particles 
which compare reasonably well with experimental data \cite{Spieles97}. 

One specific feature of these calculations is the prediction
that the system quickly enters the strangeness sector leaving the
usual $\mu_{\rm q}-T$ plane of the phase diagram. While the net-baryon
number $A_B$ decreases during the hadronization process, the s quark
chemical potential $\mu_{\rm s}$ increases from $\mu_{\rm s}=0\,{\rm
MeV}$ (for $f_s^{\rm ini}=0$ if we start with a pure u,d quark phase)
to several tens of MeV. Consequently, the strangeness fraction 
$f_s=(N_s-N_{\bar{s}})/A_{\rm tot}$ 
increases. This is the  so-called "distillation process"
\cite{CGreiner87} which can result
 in the formation of stable or metastable blobs of 
strange matter in the case of low bag model constants $B^{1/4}<180\,{\rm MeV}$. 
(In this case, a cooling of the system is predicted.) Strange and anti-strange quarks, 
produced in equal amount in the
hot plasma, do not hadronize at the same time. 
Because of the excess of
massless (u and d) quarks (as compared to their anti-quarks) in the case
of a baryon rich plasma, the hadronization of $K^+$ and
$K^0$ (containing {$\bar s$} quarks) will be favored preferentially to
$K^-$ and {$\bar K_0$} ( containing s quarks). 
The evaporation from the surface of the hadron gas, 
which is rich of anti-strangeness, 
carries away $K^+$, $K^0$,... thus charging up the 
remaining mixed system with net 
positive strangeness. 
Fig.1 shows, for two rather different  bag constants, 
the time evolution \cite{Spieles97} of the baryon
number $A$, the strangeness fraction $f_s$ and the temperature $T$ 
for initial entropy $S/A_{\rm ini}=10,45$ and $f_s=0$.
Another prediction of interest  is the hadronic freeze-out
time. Strange and antistrange hadrons are not emitted at the same time
since the hadron densities in the outer layer are dictated by the
strongly time dependent chemical potentials and the time dependent
temperature \cite{CGreiner91,Spieles97}. 
One should also notice that the difference
between $K^+\,N$ and $K^-\,N$ cross sections in a baryon
rich gas will slow down the diffusion of negative strangeness (s),
thus leading to an additional creation of time differences
\cite{CGreiner91,Spieles97}.

In the present work, this time separation 
characterizes the transient existence of the plasma. 
In the following, we will show that it can be used to 
search for the distillation process 
by using correlation techniques of unlike particle pairs 
\cite{review,Pochodzalla87,Lednicky82}.

\section{Unequal particle Correlations: Time-ordering Sensitivity}

Particle correlations at small relative momentum are mainly driven 
by strong and Coulomb final state interactions (in the case of unlike particle pairs). 

At high bombarding energies, this effect has been often treated with a well
known size-independent Coulomb correction factor. 
Contrary, at low energies, the Coulomb effect becomes the main tool 
for the study of this evolution due to larger time sequence of emission. 

Unlike particle correlations have been used experimentally 
and theoretically for more than 15 years \cite{Lednicky82,review}, 
particularly for the study of heavy 
ion collision mechanisms around the Fermi energy 
\cite{review,Pochodzalla87,Pochodzalla8586,Erazmus95}. 
Particularly, evidence for three-body Coulomb effects on 
two-particle correlations was observed in \cite{Pochodzalla8586} 
and taken into account to describe a variety of 
two-particle correlation patterns using classical trajectory 
calculations \cite{Erazmus95} or a complete three-body quantum approach 
\cite{Led9422,Martin96} in the adiabatic limit. 
The sensitivity of unlike particle correlations 
to the order of particle emission was pointed out and studied 
using both quantum \cite{Led9422,Led96} and classical 
trajectory \cite{Gel94} approaches. 
In the framework of the former one and under the assumption of 
sufficiently small density in phase space,
the two--particle correlation function at a given relative c.m.s.
momentum $\vec{q}=2\vec{k}$ is determined by the modulus squared
of the two--particle amplitude averaged over the relative c.m.s.
coordinates $\vec{r}^*$ of the emission points. 
The sensitivity of the correlation function to the  
time delays, or generally to the space-time asymmetries 
in particle production, appears due to the dependence of the 
two-particle amplitude on the scalar product $\vec{k}\vec{r}^*$. 
In particular, in the limit of large relative emission times 
$t=t_1-t_2$, $v|t|\gg r$, the Lorentz transformation from 
the source rest frame to the two-particle c.m.s ($v$ and $\gamma$ are
the pair velocity and Lorentz factor): 

\begin{equation}
r_L^*=\gamma (r_L-vt),\hspace{0.7cm} r_T^*=r_T, 
\end{equation}

indicates that the vector $\vec{r}^* \approx - \gamma \vec{v} t$ 
is only slightly modified by
averaging over the spatial location of the emission points in the
rest frame of the source. So the vector $\vec{r}^*$ is nearly parallel 
or antiparallel to the pair velocity $\vec{v}$, depending on the sign
of the time difference $t$. The dependence of the amplitude on 
$\vec{k}\cdot\vec{r}^*$ is thus transformed into the dependence of the
correlation function on $-\vec{k}\cdot\vec{v}\gamma t$.
Therefore, the sensitivity to the sign of the time difference $t$
is due to the odd part, in $\vec{k}\cdot\vec{v}$, of the correlation
function. The mean relative emission time, including its sign, can be
determined by comparing the correlation functions
$R^+$ and $R^-$ corresponding to $\vec{k}\cdot\vec{v}>0$ and $<0$
\cite{Led96}.
Noting that for particles of equal masses the sign of 
$\vec{k}\cdot\vec{v}$ coincides with the sign of the velocity 
difference $v_1-v_2$, we can see the simple classical meaning of the 
above selection. It corresponds to the intuitive expectation of 
different particle interaction in the case when the faster particle is 
emitted earlier as compared to the case of its later emission. 

For charged particles characterized by a large Bohr radius, 
$|a|\gg \langle r^* \rangle, |{\rm Re}f|$ 
($f$ is the amplitude due to the strong interaction), the ratio 
$(1+R^+)/(1+R^-)$ at very small $q$ takes on a simple 
analytical form \cite{Lednicky98}: 
\begin{equation}
 (1+R^+)/(1+R^-) \approx 1+2\langle r_L^* \rangle /a \rightarrow 
  1-2 \langle \gamma v t \rangle /a, 
\end{equation}
where the arrow indicates the limit $v|t| \gg r$.
Thus, for $K^+K^-$-system ($a=-111$ fm), each fm in the 
asymmetry $\langle r_L^* \rangle$ or $\langle \gamma v t \rangle$ 
transforms in a 2$\%$ change of the $(1+R^+)/(1+R^-)$ ratio at 
$q\rightarrow 0$. 
The correlation functions $R^+$ and $R^-$ as predicted in \cite{Led9422,Led96} 
well agree with the proton--deuteron correlation
functions measured in low energy nuclear collisions
studied at GANIL \cite{Ghi93,Era96,Gou96,Nouais97}. 
As proposed in \cite{Led96}, this technique can be extended for a study of the emission time
differences between any kinds of interacting unlike particles. 

In preliminary studies using simple event generators, it has been 
demonstrated \cite{Era95,Era96,Soff97a} that $K^+ K^-$ pairs with mean emission time
differences as small as $\pm 5\, {\rm fm/c}$ can give rise to observable
differences between $R^+$ and $R^-$ correlation functions. It has been also
found that most of the effect at $q < 10\,{\rm MeV/c}$ originates from the 
Coulomb interaction between the two kaons and that, in full correspondence
with Eq. (2), the later (on the average) emission of $K^-$'s is
associated with the $(1+R^+)/(1+R^-)$ ratio less than unity.

In this paper, we will use this method in conjunction with the dynamical
two phase description (see previous section) in order to quantitatively
demonstrate its sensitivity to the predicted delays between the $K^+$ and $K^-$
emission related to the production of a transient strange quark matter state.

Three sets of parameters \cite{Soff97a,Spieles97} have been
chosen:\\
{\it set 1 $\&$ 2} (representing Au+Au at AGS energies \cite{Spieles97}):\\
initial mass $A= 394$, entropy per baryon $S/A= 10$, initial net
strangeness $f_s= 0$;\\
{\it set 1}: $B^{1/4} = 160\,{\rm MeV}$, strangelets (albeit metastable) are formed \cite{Spieles97}.\\
{\it set 2}: $B^{1/4} = 235\,{\rm MeV}$, strangelets are not distilled \cite{Spieles97}. 
They are not (meta-)stable.\\
{\it set 3} ( representing S+Au at SPS energies \cite{Spieles97}):\\
bag model constant $B^{1/4} = 235\,{\rm MeV}$, 
initial mass $A= 100$, entropy per baryon $S/A= 45$, initial net
strangeness $f_s= 0$.\\

In the case of the low bag constant ($B^{1/4}=160\,{\rm MeV}$) - set 1 -
a cooling of the system is predicted leading to rather long
kaon emission times. In the early stage mainly $K^+$'s are emitted
(see Fig. 1, l.h.s. column) so that a cold strangelet emerges
in a few tens fm/c.
According to Eqs. (1), (2) and the corresponding discussion, the
later emission of $K^-$'s should lead to the correlation function ratio
$(1+R^+)/(1+R^-)$ less than unity provided that the asymmetry 
$\langle r_L^* \rangle$
in $K^+K^-$ c.m.s. is dominated by the time asymmetry term 
$\langle vt \rangle$.
As seen from table 1 and Fig. 2 this is indeed the case: 
the spatial asymmetry term $\langle r_L \rangle$ contributes 
by less than $10 \%$ (in the same direction); 
the predicted value of the deviation
of the correlation function ratio from unity of $-26\%$ at $q \rightarrow 0$
is in good agreement with the intercept value of the calculated
ratio curve with the ordinate axis. Also, the deviation 
from unity remains important and statistically meaningful up to the region 
$q\approx30 \,{\rm MeV/c}$. Thus, in spite of experimental 
difficulties to measure correlations down to a few ${\rm MeV/c}$, 
the signal proposed for consideration is accessible to 
usual correlation measurements. 
Since for various two-particle systems produced in the mid-rapidity
region, the ordinary dynamical mechanisms lead to the intercept values
deviating from unity by less than $10\%$ \cite{Lednicky98}, we can consider the 
eventual observation of a large negative asymmetry in the
$K^+K^-$ correlation function ratio as a signal
of the strangelet formation.

In the case of the high bag constant ($B^{1/4}=235\,{\rm MeV}$) - set 2 and 3 -
the system heats up slightly.
This results in a fast hadronization \cite{CGreiner91} and, 
subsequently, in a small
difference between the $K^+$ and $K^-$ emission times \cite{Spieles97}
(see the middle and r.h.s. columns in Fig. 1).
Similar to the previous case the $K^+$'s are emitted earlier than $K^-$'s.
Also the asymmetry $\langle r_L^* \rangle$ is dominated by
the time asymmetry term $\langle vt \rangle$, the spatial one, $\langle r_L \rangle$, 
contributing by about $25\%$ in the same direction (see table 1, rows 2 and 3).
However, the asymmetry effect is now much weaker (Fig.2):
the predicted intercepts with the ordinate axis of the correlation
function ratio deviate
from unity by $-4.7\%$ and $-2.3\%$ for sets 2 and 3, 
respectively.

Clearly, in this case, such a relatively weak asymmetry effect cannot be
considered as a signal of the strangeness distillation without
detailed studies of the impact of the ordinary mechanisms 
(rescattering effects, for example).

Regarding the width of the effect seen in the correlation function
ratios (Fig. 2), it decreases with the increasing $\langle r^* \rangle$ (see table 1)
in correspondence with the narrowing of the correlation effect itself. 
Thus, for set 1 ($\langle r^* \rangle \approx 18\,{\rm fm}$) 
the effect rapidly vanishes with increasing $q$ and extends up to about 
$30 \,{\rm MeV/c}$ only. 
For sets 2 and 3 ($\langle r^* \rangle \approx 6\,{\rm fm}$ in both cases) 
the deviation from unity is softer in the same range of $q$ values and 
extends up to about $60 \,{\rm MeV/c}$.

\section{Conclusions}

We have presented a novel method which can be applied to characterize
the possible existence of a strange quark matter distillation process 
in ultra-relativistic heavy-ion collisions.
The method is based on the predictions 
\cite{Lednicky82,Led9422,Led96,Era95,Era96}
for unequal particle correlations and exploits the predicted
properties of the transient,
strange quark matter state \cite{CGreiner87,CGreiner91,Spieles96,Spieles97}, 
even if it decays on strong interaction time scales. Using the
description of this strangeness distillation process by a dynamical
evolution model for the mixed QGP-hadronic 
phase, we have quantitatively demonstrated the
sensitivity to the bag constant correlated to the stability 
of the quark matter possibly
encountered in available experimental situations.
 
For the lower bag constant case a strong and sharp negative asymmetry
effect is predicted in the correlation function ratio, thus offering a clear signal
of the predicted strangeness distillation process provided the experimental 
resolution and statistics are sufficient to measure the ratio down to about 
$10\,{\rm MeV/c}$. 
Weaker though wider signals are predicted for the higher bag constant hypothesis.
In this case the competition with other collision mechanisms generating
the asymmetry is possible. These mechanisms, including the expansion of
the hadronic system, are currently under investigation [32,33]. 

Finally,we think that the
unlike particle correlation technique may  offer a very valuable tool
to disentangle between different and presently debated scenarios for the
phase transition in QCD matter at finite net baryon density.

\section{Acknowledgements}
The authors would like to thank B.~Erazmus, L.~Martin, D.~Nouais, 
C.~Roy, and A.Dumitru for useful discussions and/or for providing  
numerical programs.
D.~A. is pleased to thank the Institut f\"ur Theoretische Physik 
at the University of Frankfurt for invitation and kind 
hospitality.

\newpage
\section*{References}



\newpage

\begin{tabular}{|c||r|r|r|r|}\hline
set & $\langle vt \rangle$ (fm) & $\langle r_L \rangle$ (fm) & $\langle r_L^* \rangle$ (fm) & $\langle r^* \rangle$ (fm)  \\ \hline \hline
1 & -10.9  & 1.1& 14.3 & 17.5 \\ \hline
2 &  -1.4 & 0.5 & 2.6 & 5.5  \\ \hline
3 &  -0.7 & 0.2 & 1.3 & 6.3  \\ \hline
\end{tabular}

\begin{table}
\caption{Mean values of the relative space-time coordinates for pairs with 
relative momenta $q \le 50 {\rm MeV/c}$  
calculated within the mixed phase thermodynamical model (see text).}
\end{table}


\newpage

\begin{figure}
\caption{Time evolution of baryon number $A$, strangeness 
fraction $f_s$, temperature $T$ and yields of $K^+$ and $K^-$ 
for {\it set} 1 (left column), {\it set} 2 (central column)
and {\it set} 3 (right column) (see text).} 
\end{figure}

\begin{figure}
\caption{Ratios of correlation functions $(1+R^+)/(1+R^-)$ 
for three parameter sets in a mixed-phase thermodynamical model (see text). 
The strange quark matter distillation process predicted in the case of a 
low bag constant (upper curve) results in a measurable deviation 
from unity in the range $q<30\,{\rm MeV/c}$.}
\end{figure}

\newpage

\psfig{figure=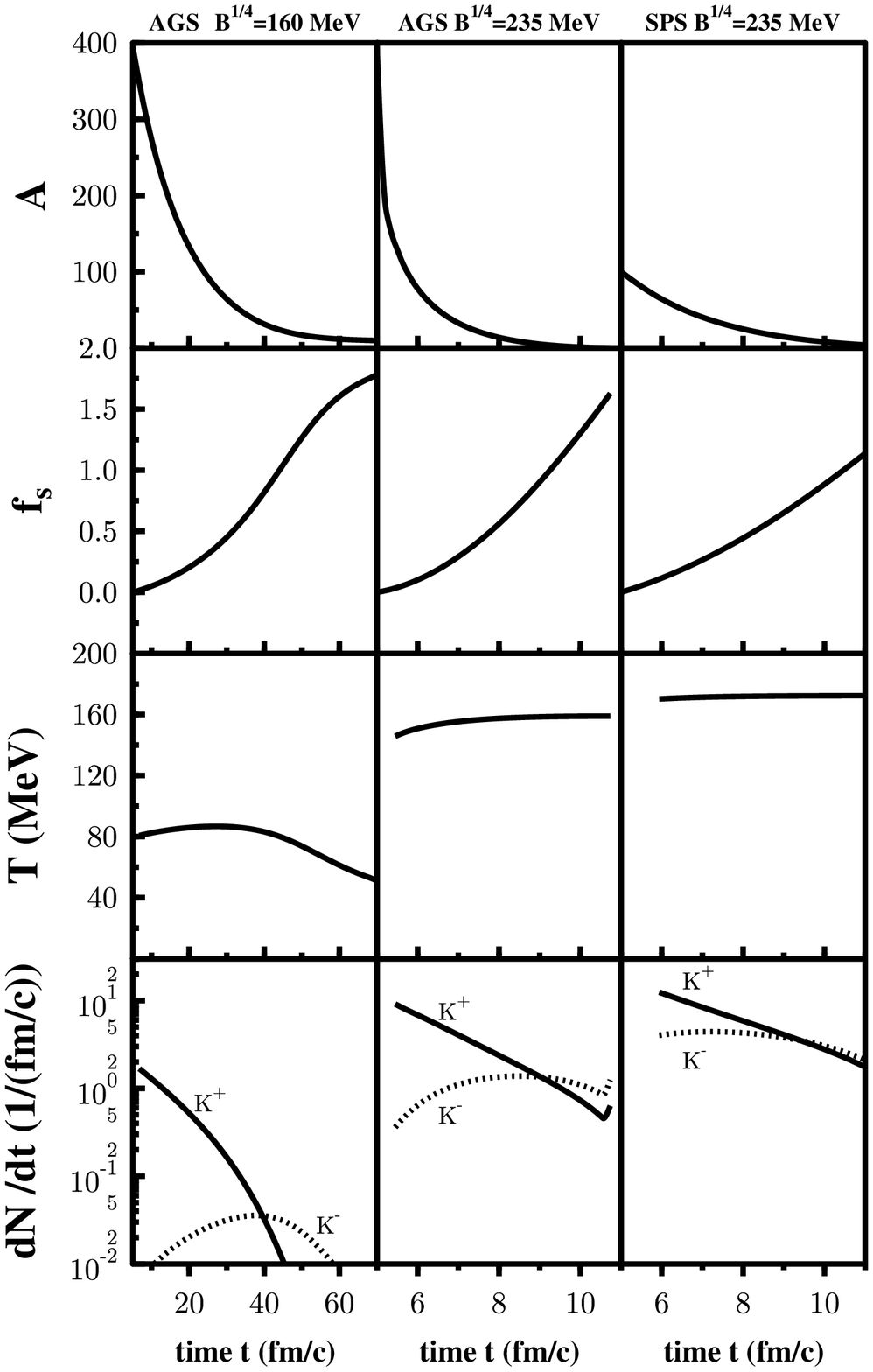}

\newpage
\psfig{figure=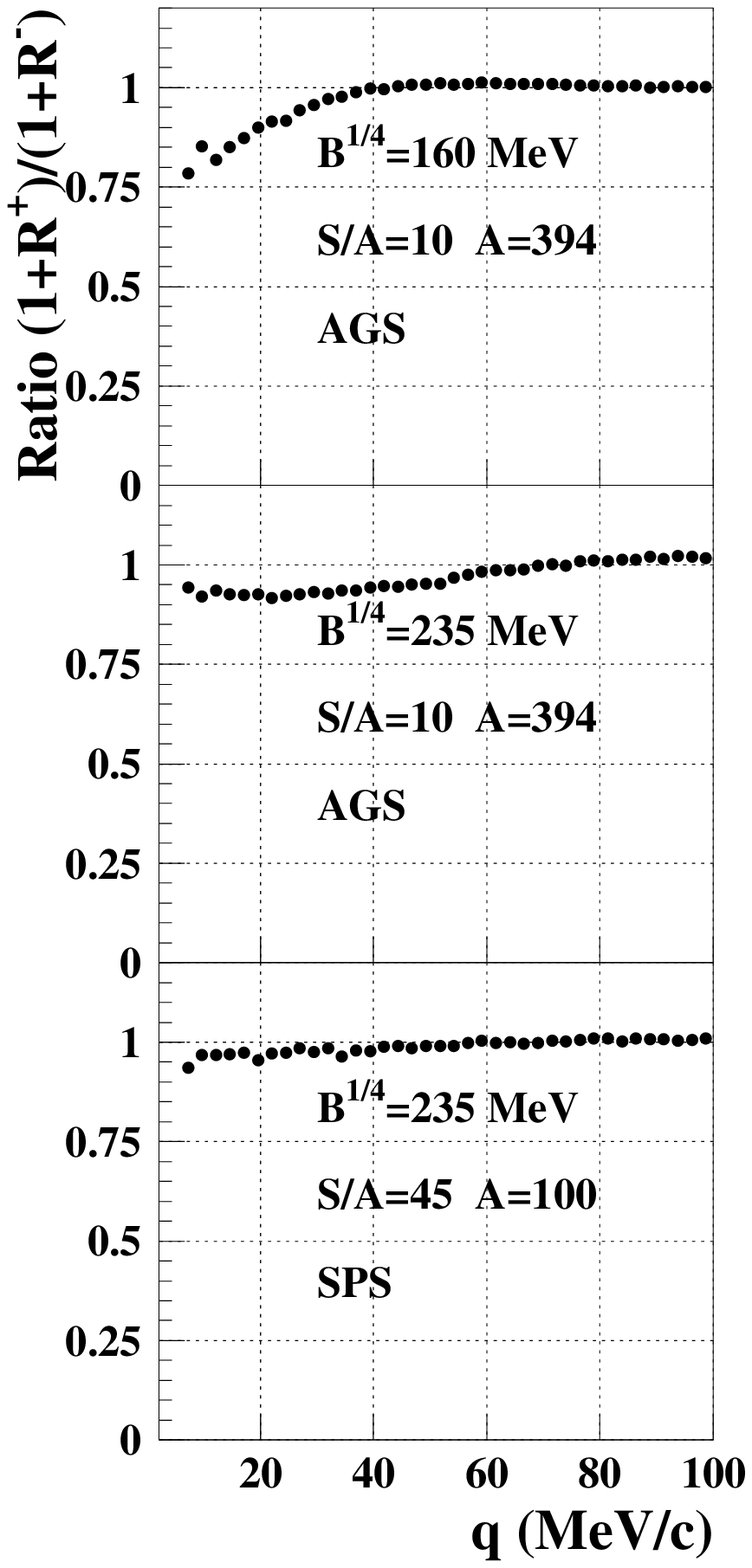,height=18cm,width=10cm}
\vspace*{1cm}
\center{\large \bf Fig.2}

\enddocument